\documentclass{ws-procs975x65}

\usepackage{graphicx}

\DeclareMathOperator{\Order}{\mathcal{O}}
\allowdisplaybreaks

\begin{document}

\title{\uppercase{Influence of internal structure on the motion of test bodies in extreme mass ratio situations}}

\author{\uppercase{Jan Steinhoff}}
\address{CENTRA, Instituto Superior T\'ecnico, Avenida Rovisco Pais 1, 1049-001 Lisboa, Portugal
\\jan.steinhoff@ist.utl.pt}

\author{\uppercase{Dirk Puetzfeld}}
\address{ZARM, University of Bremen, Am Fallturm, 28359 Bremen, Germany
\\dirk.puetzfeld@zarm.uni-bremen.de}

\begin{abstract}
We present some recent results on the motion of test bodies with internal structure in General Relativity. On the basis of a multipolar approximation scheme, we study the motion of extended test bodies endowed with an explicit model for the quadrupole. The model is inspired by effective actions recently proposed in the context of the post-Newtonian approximation, including spin-squared and tidal contributions. In the equatorial plane of the Kerr geometry, the motion can be characterized by an effective potential of the binding energy. We compare our findings to recent results for the conservative part of the self-force in astrophysically realistic situations.
\end{abstract}

\keywords{Approximation methods; Equations of motion; Extreme mass ratios}

\bodymatter\bigskip

One interesting prospect of upcoming gravitational wave astronomy is the ability to probe the internal structure of compact astrophysical objects. Such objects can be modeled by different methods in the context of General Relativity. Here we employ a multipolar approximation scheme \cite{Dixon:1979,Steinhoff:Puetzfeld:2009:1}, in which the equations of motion for test bodies with internal structure take the following form up to the quadrupolar order:
\begin{eqnarray}
\frac{\delta p_a}{d s} &=& \frac{1}{2} R_{abcd} u^{b} S^{cd} + \frac{1}{6} \nabla_a R_{bcde} J^{bcde}, \\
\frac{\delta S^{ab}}{d s} &=& 2 p^{[a} u^{b]} - \frac{4}{3} R^{[a}{}_{cde} J^{b]cde},
\end{eqnarray}
where $s$ is the proper time, $u^a$ is the 4-velocity, $p_a$ the 4-momentum, $S^{ab}=-S^{ba}$ the 4-spin, and $J^{abcd}$ is the quadrupole moment with $J^{abcd} = J^{[ab][cd]} = J^{cdab}$ and $J^{[abc]d} = 0$. The spin contributions were already discovered by Matthisson\cite{Mathisson:1937} and later on by Papapetrou\cite{Papapetrou:1951:3}. Note that there is no dynamic equation for the quadrupole (or higher multipoles). Such a quadrupole model -- as well as a supplementary condition for the spin -- must be added by hand in order to close the system of equations, and thus allow for a unique prediction of the motion. See also Ref.~\citen{Steinhoff:Puetzfeld:2009:1} for a discussion of different supplementary conditions and conserved quantities at different multipolar orders.

Following Ref.~\citen{Steinhoff:Puetzfeld:2012}, we adopt a quadrupole model for astrophysically realistic scenarios, which was recently developed within a post-Newtonian context, i.e.\
\begin{eqnarray}
 J^{abcd} &=& - \frac{m}{\underline{m}} \left[ \frac{1}{\underline{m}} p^{[a} Q^{b]cd} + \frac{1}{\underline{m}} p^{[d} Q^{c]ba}	+ \frac{3}{\underline{m}^2} p^{[a} Q^{b][c} p^{d]} \right], \label{quadmodel} \\
 Q^{ab} &=& c_{ES^2} S^a{}_{e} S^{be} - \mu_2 E^{ab} , \quad \quad Q^{bcd} = - \frac{2 \sigma_2}{\underline{m}} \eta^{dc}{}_{ea} p^e B^{ba}, 
\end{eqnarray}
where $Q^{ab}$ is the mass quadrupole and $Q^{bcd}$ the flow quadrupole\cite{Ehlers:Rudolph:1977}. The mass-like quantities are defined by $\underline{m}^2 := p^a p_a$ and $m := p_a u^a$. The quantities $c_{ES^2}$, $\mu_2$, and $\sigma_2$ are assumed to be constants, and parameterize quadrupole deformations induced by the spin and by tidal forces of the spacetime. Furthermore,  $E^{ab}=\frac{1}{\underline{m}^2} R_{acbd} p^c p^d$ represents the gravito-electric tidal field, and $B^{ab}=\frac{1}{2 \underline{m}^2} \eta_{aecd} R_{bf}{}^{cd} p^e p^f$ the gravito-magnetic (frame-dragging) tidal field, where $R_{abcd}$ is the Riemann tensor and $\eta_{abcd}$ the volume form. The quadrupole model in (\ref{quadmodel}) can be derived from effective actions\cite{Porto:Rothstein:2008:2, Goldberger:Rothstein:2006:2, Damour:Nagar:2009}. 

Given this quadrupole model, the motion of a mass-spin-quadrupole test body in the equatorial plane of the Kerr geometry is studied. We further assume that the spin of the test body is aligned with the rotation axis of the background spacetime. In the absence of a quadrupole, this problem can be solved in a simple manner\cite{Rasband:1973}, and we have shown that this method is in fact still applicable for the considered quadrupole model\cite{Steinhoff:Puetzfeld:2012}. The method makes use of the fact that the conserved quantities, the spin supplementary condition, and the constraints on the orbital configuration are enough to uniquely fix the 10 dynamic variables contained in $p_a$ and $S^{ab}$.

The spin supplementary condition ($S^{ab} p_b = 0$) contains three independent equations, while the constraint on the orbit provides three further independent conditions, one due to equatorial orbits ($p^{\theta}=0$) and two due to spin alignment ($S^{a\theta} = 0$). So we need to identify $10-3-3=4$ conserved quantities in order to solve for $p_a$ and $S^{ab}$ algebraically. The Killing vectors of Kerr spacetime $\partial_t$ and $\partial_{\theta}$ give rise to conserved energy $E:=E_{\partial_t}$, and total angular momentum $J:=E_{-\partial_{\theta}}$, where\cite{Ehlers:Rudolph:1977} $E_{\xi}:= p_a \xi^a + \frac{1}{2} S^{ab} \nabla_a \xi_b$. Furthermore, it can be shown\cite{Steinhoff:Puetzfeld:2012} that the spin-length $S:= \sqrt{ \frac{1}{2} S_{ab} S^{ab} }$ is conserved for the considered quadrupole model. The identification of a conserved mass-like quantity is most problematic, as it crucially depends on the adopted quadrupole model. We were only able to identify a mass-like quantity that is conserved in an approximate sense. If we utilize a multipole counting scheme of the form
\begin{equation}
\mu = \Order{(\epsilon^0)} = p^a , \quad
\frac{\delta p^a}{d s} = \Order{(\epsilon^1)} = S^{ab} , \quad
\frac{\delta S^{ab}}{d s} = \Order{(\epsilon^2)} = J^{abcd},
\end{equation}
then the mass-like quantity $\mu$,
\begin{equation}
\mu := \underline{m} + \frac{c_{ES^2}}{2} E_{ab} S^a{}_{c} S^{cb} + \frac{\mu_2}{4} E_{ab} E^{ab} + \frac{2\sigma_2}{3} B_{ab} B^{ab} , \label{const_mass}
\end{equation}
is conserved up to the order $\Order{(\epsilon^3)}$. This formula for $\mu$ can in fact be easily derived from the underlying effective action.

Now we are in a position to solve for $p_a$ and $S^{ab}$. Most important is the equation for $p^r$. It turns out, that  $(p^r)^2$ is equal to a polynomial of second order in $E$. With the roots of this polynomial denoted by $U_{+}$ and $U_{-}$, one can write
\begin{equation}
\left(p^r\right)^2 \propto (E - U_{+}) (E - U_{-}) .
\end{equation}
For $p^r$ to be a real number we need to have both $E \leq U_{+}$ and $E \leq U_{-}$, or both $E \geq U_{+}$ and $E \geq U_{-}$. It turns out that the important relation is just $E \geq U_{+}$ for the most relevant part of the parameter space. This justifies to call $U_{+}$ effective potential. The test body can only move in the region where $E \geq U_{+}$ and the turning points are given by $E = U_{+}$, because then $p^r = 0$ -- which implies $u^r = 0$. Therefore the minimum of $U_{+}$ defines circular orbits.

We compared the binding energy $E$ for circular orbits with recent results for the conservative part of the self-force\cite{LeTiec:Barausse:Buonanno:2012} and with various post-Newtonian Hamiltonians\cite{Steinhoff:Puetzfeld:2012}. The former is illustrated for the astrophysically realistic case of a very rapidly rotating (small) black hole in a Schwarzschild background in Fig.~1. The mass ratio is formally $q=1$, so the curves must be scaled to a realistic case ($q \lesssim 10^{-2}$). Self-force and linear spin effects scale as $\propto q$, the others as $\propto q^2$. In a Kerr background the last stable circular orbit can be very close to the horizon, such that the discussed effects can be some orders of magnitude stronger. A more complete discussion including the tidal quadrupole contributions proportional to $\mu_2$ and $\sigma_2$ can be found in Ref.~\citen{Steinhoff:Puetzfeld:2012}.
\begin{figure}[t]
\begin{tabular}{m{8.2cm} m{3.5cm}}
\includegraphics{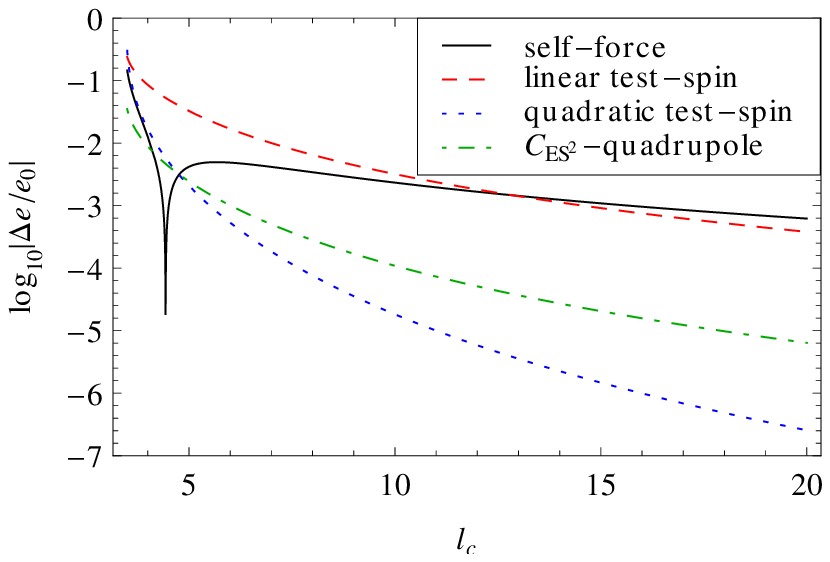}
\vspace*{-0.8cm}
&
Fig.~1~
Various corrections to the binding energy $e$ for a maximally spinning (small) black hole in a Schwarzschild background. $l_c$ is the \emph{orbital} angular momentum. Curves must be scaled to the actual mass ratio, see text.
\end{tabular}
\end{figure}

\emph{Acknowledgments} ---
This work was supported by the Deutsche Forschungsgemeinschaft (DFG) through the grants STE 2017/1-1 (J.S.) and LA-905/8-1 (D.P.); and in an initial phase through the SFB/TR7 (J.S.).

\end{document}